\documentstyle [prl,aps,multicol,epsfig]{revtex}

\begin{document}

\bibliographystyle{prsty}

\draft

\title{Anisotropic Disorder in High Mobility 2D Heterostructures and its 
Correlation to Electron Transport}

\author{R.L. Willett, J.W.P. Hsu, D. Natelson, K.W. West, L.N. Pfeiffer}

\address{Bell Laboratories, Lucent Technologies, Murray Hill, NJ 07974}

\maketitle

\begin{abstract} 
Surface morphology of high mobility heterostructures is examined and
correlated with d.c. transport.  {\it All} samples examined show
evidence of lines in the [1$\bar{1}$0] direction with 
roughness ranging from small amplitude features to severe anisotropic
ridges. Transport in these samples is consistent with that in samples
having artificially induced 1D charge modulations. The native surface
properties reflect a prevalent, anisotropic disorder affecting 
2D electron conduction. Importantly, the native lines are orthogonal
to the stripes theoretically proposed to explain high Landau level
transport anisotropies.

\end{abstract}

\begin{multicols}{2}

Correlated electron effects in 2D heterstructures can be observed
because the disorder of these structures has been understood and
subsequently controlled to facilitate low energy electronic
interactions. The fractional quantum Hall effect(FQHE)\cite{Tsui:82}
was observed in modulation doped heterostructures\cite {Stormer:79} as
this specific device provides spatial separation of ionized impurities
and the interacting 2D electron system (2DES).  Molecular beam epitaxy
(MBE) has produced extraordinarily pure systems\cite{Pfeiffer:88} with
electron mobilities exceeding 10$^{7}$ cm$^{2}$/V-sec.  Such
structures exhibit phenomena ranging from composite fermion Fermi
surface effects\cite{Olle:98} at $\nu $=1/2, to transport anisotropies
at high Landau levels\cite {Du:99,Lilly:99}, proposed to be electron
stripe and bubble phases\cite{Koulakov:96}. In the composite fermion
system, density variations play a particularly significant role as
disorder due to the association of charge and fictitious magnetic
field\cite{Willett:97}.  Establishing sources of lateral charge
variation within the high mobility interface is therefore important
for understanding disorder influencing the correlated 2DES.

The surface of high mobility heterostructures provides an accessible
plane to study possible density related disorder.
Surface morphology can vary under different MBE growth
conditions, and growth anisotropies can occur along certain
crystallographic directions\cite{Johnson:94,Alexandre:85}. The length
scales of these variations are large, on the order of
microns, and so are imminently measureable. Key questions are then, do
these large-scale imperfections occur in high mobility
heterostructures, and do they influence the electronic properties of
the 2D layer?

To answer this we have examined surface morphology using atomic force
microscopy and studied transport of a series of high mobility
heterostructures. In an extensive pool of high mobility wafers we
observed a range of surface properties related to known crystal growth
instabilities. Our principal finding is that {\it all} surfaces are
observed to have some formation of ridges along the [1$\bar{1}$0]
direction. These lines exist within a range of large-scale topographic
fluctuations that vary from nearly isotropic mounds to stark
anisotropic ridges consistently along the [1$\bar{1}$0]
direction. Transport in samples with severe surface roughness is shown
to be similar to that observed in heterostructures with artificially
induced 1D charge density modulations\cite{Willett:97,Smet:98}.  In
samples with the less severe native surface roughness, transport
anisotropies can be identified in the lowest Landau level which are
consistent with the surface morphology.

A focus of this study is the surface morphology of ultra-high mobility
samples that demonstrate both extensive FQHE features and high Landau
level transport anisotropies.  These high quality samples show i) FQHE 
states manifest properties associated with the anisotropic growth, and
ii) the native surface lines, which we associate with charge modulation
as shown by transport, are oriented {\it orthogonally} to the stripes 
proposed in the theoretical models\cite{MacDonald:00,Koulakov:96} of 
high Landau level anisotropies.  

The MBE-produced wafers in this study were all grown on [001] oriented
substrates with a buffer layer of minimally 1~$\mu$m thick GaAs or
AlGaAs/GaAs superlattice.  Transport properties and surface morphology
were independent of whether heterostructures are single interfaces or
quantum wells.  The 2D layer resides less than 300~nm below the
surface for all samples studied.  The wafers focused upon here
(and shown in Figures 1a-c) were produced under essentially the
same conditions, indicating that the properties determining surface
morphology are not well understood.

A large number of heterstructure wafers with mobilities greater than
8$\times$10$^{6}$cm$^{2}$/V-sec were examined, and shown here are
samples which are representative of the full range of surface
morphologies. Light microscopy was also used to examine
heterostructure surfaces; using phase contrast optics, most of the
larger surface features can be detected readily at relatively low
magnifications.

Surface morphology was examined using atomic force microscopy
operating in tapping mode.  Topographic images (Figure \ref{fig:Figure
1}, right column) depict surface height variations.  The change in
cantilever oscillation amplitude, proportional to the
derivative of topographic changes, is also shown (Fig. \ref{fig:Figure
1}, left column).  Transport was performed using standard lock-in
techniques at low frequencies.

\begin{figure}
\epsfclipon
\epsfxsize=8.5cm
\epsfbox{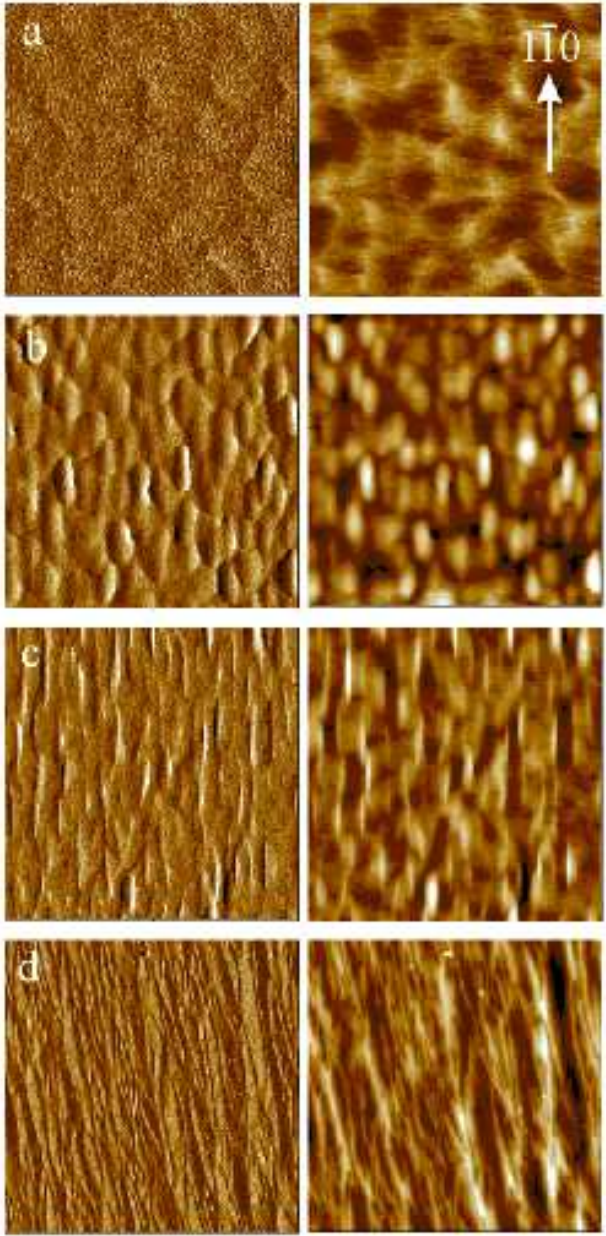}
\caption{Representative range of surface morphologies of high mobility
heterostructures as measured with atomic force microscopy.  Each field
size is 20~$\mu$m $\times$ 20~$\mu$m .  The images on the right are surface
topography, and the images on the left show variations in cantilever
amplitude which are proportional to the topography derivative.  Sample
mobilities and total vertical range for the topographic images (black
to white) are (a)16x10$^{6}$ cm$^{2}$/V-sec and 40~\AA, (b)
27x10$^{6}$ cm$^{2}$/V-sec and 100~\AA, (c)23x10$^{6}$ cm$^{2}$/V-sec
and 100~\AA, (d) 8x10$^{6}$ cm$^{2}$/V-sec and 300~\AA.}
\label{fig:Figure 1}
\end{figure}

Figure \ref{fig:Figure 1} shows surface morphology for
heterostructures ranging from smoothest (a) to roughest (d). All
samples show features, lines, along the [1$\bar{1}$0] direction. In
the lowest overall amplitude surface topography, Figure
\ref{fig:Figure 1}(a), the dominant morphology is a coarsely isotropic
``orange peel'' effect with a diameter of about 4 $\mu $m and height
range of less than 40 \AA . Within these large features, fine lines
corresponding to atomic scale terracing are observed alligned in the
[1$\bar{1}$0] direction. This predilection toward line formation in
the [1$\bar{1}$0] direction becomes more apparent as the overall
surface roughness increases. In Figure \ref{fig:Figure 1}(b) the
orange peel texture has evolved to a surface with mounds clearly
oriented in the [1$\bar{1}$0] direction, and with vertical excursion
of almost 100 \AA .  Figure \ref{fig:Figure 1}(c) shows another
representative sample in the range of common surface morphologies.
Severe ridges are present with about the same period as that shown in
Figure \ref{fig:Figure 1}(b) and with almost the same vertical
excursion. It is important to note that wafers generally display
morphology ranging from that shown in (a) to that shown in (c).  The
final AFM image (Figure \ref{fig:Figure 1}d) shows an extreme case of
surface roughness: this was induced by growing a particulary thick
buffer layer of about 10~$\mu$m, capped by a standard interface. The
vertical excursion is large, up to 300 \AA , with line formation
distinctly in the [1$\bar{1}$0] direction.  Note that ridges can
traverse the full field of the scan.

\begin{figure}
\epsfclipon
\epsfxsize=8.5cm
\epsfbox{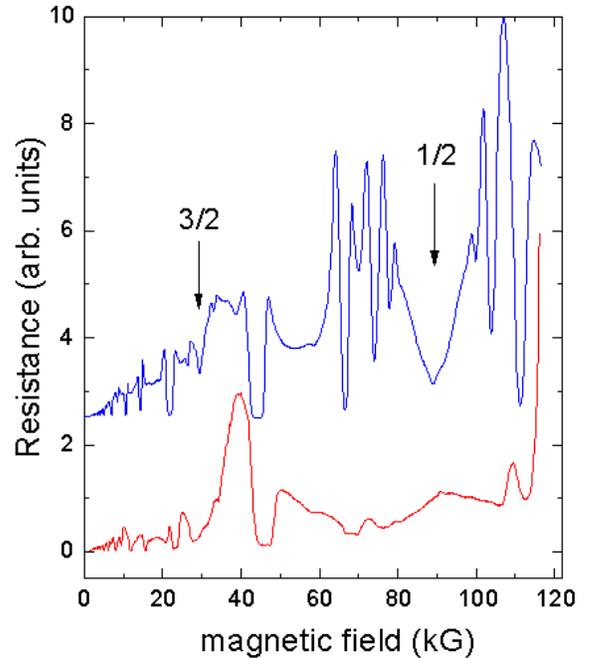}
\caption{Magnetotransport measured for the sample of Figure 1d.  The red 
trace corresponds to current driven in the [1$\bar{1}$0] direction and the 
blue trace is current driven in the [110] direction.  Temperature=290~mK.}
\label{fig:Figure 2}
\end{figure}

The common property of the above described range of morphologies is
the formation of lines in the [1$\bar{1}$0] direction. This effect is
due to a previously recognized MBE growth instability
\cite{Johnson:94} with fast growth rate along the direction of the As
dangling bonds\cite{Fleming:80}.  Extensive growth or lowered
temperatures during growth can accentuate this effect.

A second mechanism contributing to this formation of lines in the
[1$\bar{1}$0] direction is from the AlGaAs layering. A temperature and
Al concentration dependent growth preference in the [1$\bar{1}$0]
direction has been documented\cite{Alexandre:85}. It provides a
smaller length scale ridge formation than the above described
mechanism, and is at work in these samples due to the presence of
GaAs/AlGaAs layering.  Suppressing this contribution to ridge
formation involves growth at the appropriate temperature and Al
concentration; suppression of the previously noted growth instability
is more difficult.

Transport results show that the surface morphology is associated with
charge variation at the 2D electron layer.  This precept is supported
by the data of Figure \ref{fig:Figure 2}: it shows typical transport
data from the sample of Figure \ref{fig:Figure 1}d having severe
natural lines, with current driven across and along the [1$\bar{1}$0]
direction. The key features in this data are the large
minimum in resistance at 1/2 filling factor for current driven across
[1$\bar{1}$0] (across the native lines), accompanied by a peak
at 1/2 for current driven along [1$\bar{1}$0].
Note also that current driven across the native lines
produces narrow FQHE states; conversely, current driven along the
lines produces relatively widened FQHE states.

\begin{figure}
\epsfclipon
\epsfxsize=8.5cm
\epsfbox{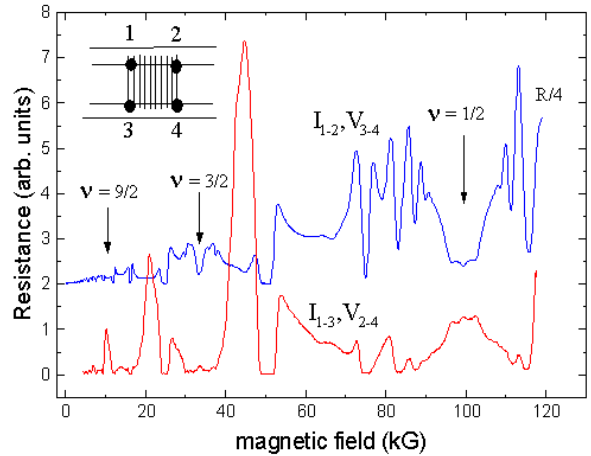}
\caption{Resistances as described by inset for 1D modulation etched into the 
heterostructure surface.  The period of tahe modulation is 1.2~$\mu$m and the 
temperature is 290~mK.}
\label{fig:Figure 3}
\end{figure}

These effects are similar to those observed in transport in the lowest
Landau level for a 1D charge density modulation artificially induced
on a heterostructure: results from such a sample are shown in Figure
3.  Artificial 1D charge density modulation is achieved by etching an
array of lines into the surface of the heterostructure.  The etch
depth can be small, typically about 150 \AA, to induce transport
consequences.  The artificially modulated system shows a peak at 1/2
for current driven along the imposed lines and a large minimum for
current driven across the lines, just as seen in the sample of Figure
\ref{fig:Figure 1}(d) with native lines.  In addition, the FQHE states
show relative narrowing when current is driven across the modulation
lines and broadening when driven along the modulation lines; again,
this is similar to the effect observed for the respective transport
directions in the native line sample of Figure \ref{fig:Figure 1}(d).
The central minimum and peak at 1/2 for the two transport directions
in the artificially modulated sample of Figure \ref{fig:Figure 3} show
small features at and symmetrically about 1/2. The lateral minima
(peaks) correspond to geometric resonance of the charged carrier
motion with the regular period of the
modulation\cite{vonoppen:98,Willett:99}, and the central minimum
(peak) at 1/2 is due to ``snake'' states\cite{Muller:92}.  These occur
as a consequence of large radius composite fermion cyclotron orbit
motion along an effective B-field gradient centered about zero
effective magnetic field, as expected to exist along the lines of a
charge density gradient. Note the presence of such snake state
transport at 1/2 in the sample with strong native lines; the geometric
resonances are absent due to the aperiodic nature of the native charge
modulation.  Other features are common to the magnetotransport spectra
of Figures \ref{fig:Figure 2} and \ref{fig:Figure 3}; for example note
the large peak between $\nu =$~3/2 and 1 in the red traces.  The
presence of these multiple specific features in both the native line
sample and the artificially modulated sample clearly show empirically
that the native surface lines in Figure \ref{fig:Figure 1}d are indeed
associated with charge modulations in the 2D electron layer.

\begin{figure}
\epsfclipon
\epsfxsize=8.5cm
\epsfbox{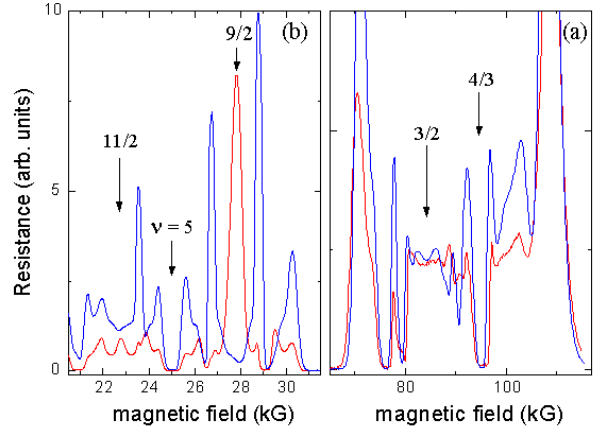}
\caption{Magnetotransport around filling factors 3/2 (a) and 9/2 (b)
from the sample shown in Figure 1(c).  
The current is driven along [110] (blue traces)
or along [1$\bar{1}$0] (red traces).
The temperature is 110~mK.  Substantially larger quantitative 
anisotropy at 9/2 is observed at lower temperatures.}
\label{fig:Figure 4}
\end{figure}

Less coarse surface effects also produce transport consequences in the
lowest Landau level.  Shown in Figure \ref{fig:Figure 4}(a) is the
magnetotransport around 3/2 for the sample pictured in Figure 1(c).
The native modulations of this sample induce transport properties
similar to those noted in the artificially charge modulated system:
the FQHE states are wider in the orientation of current along the
[1$\bar{1}$0] direction (along the lines).  This broadening of about
$\delta B/B \sim $ 0.5\% reflects a spread in the density of $\delta
n/n \sim $ 0.5\%.  This property was found in all of the large number
of samples of similar surface morphologies that were examined.
Additionally, current driven along the lines results in a less well
developed minimum at 3/2 and poorly formed high order fractions when
compared to transport in the orthogonal direction.  These specific
effects are also observed in samples with less severe surface
morphology. The sample shown in Figure \ref{fig:Figure 1}b
demonstrates similar broadening of the FQHE for current along the
[1$\bar{1}$0] direction, particularly at low temperatures ($<$
100~mK). Samples with morphology approaching that of Figure
\ref{fig:Figure 1}a can likewise demonstrate these effects at very low
temperatures.

A striking finding of our study is that samples with relatively severe
surface mophology demonstrate beautifully the transport anisotropies
in the high Landau levels that have recently been
documented\cite{Du:99,Lilly:99} and explained\cite{Koulakov:96} using
a new striped phase ground state.  Figure \ref{fig:Figure 4}(b)
shows transport in two directions from the sample of Figure
\ref{fig:Figure 1}(c), which has substantial native line formation in
the [1$\bar{1}$0] direction. Figure \ref{fig:Figure 4}(b) shows the
characteristic deep minimum at 9/2 for current driven across the
[1$\bar{1}$0] direction (across the native lines) and also shows a
peak at 9/2 for current along the native lines. Empirically this is
the same as the transport properties observed in the lowest Landau
level for the artificially modulated 2DES. {\it Most importantly, the
native lines observed in our morphology studies, which apparently act
as charge lines as supported by the sample's lowest Landau level
transport, are orthogonal to the theoretically proposed stripes used
to explain the high Landau level transport.} The sample from Figure
\ref{fig:Figure 1}b, which has less severe but still significant
anisotropy in the surface morphology shows similar high Landau level
anisotropies.

In summary, surface morphology of high mobility heterostructures has
been examined and found to show line formation in the [1$\bar{1}$0]
direction with varying degrees of severity over the full complement of
wafers tested.  These surface properties are correlated to transport
results, with even mild surface roughness associated with distinct
transport features.  These transport features are like those seen
specifically in samples with artificial 1D charge density modulations,
suggesting similar transport mechanisms are at work in both
systems.  These pervasive native lines represent an important form of
systematic, anisotropic disorder that is of particular significance
in ultra-high mobility samples, as it is here that their
consequences can be readily observed.

This finding of native density anisotropy in ultra-high mobility
samples has particular pertinence in considering transport
in high Landau levels, since the native lines are orthogonal 
to the direction of the stripes proposed theoretically to explain the transport
anisotropy.  It is tempting to consider that the same mechanism responsible for
effects observed around 1/2 in the artificial 1D charge modulated
samples is at play at 9/2 in the ultra-high mobility samples, with the
1D charge modulation provided by the native line formation.
In spite of numerous empirical similarities, however, this is unlikely 
as Fermi surface formation would be required at the high Landau level
half-filling.  The native lines may have some influence in breaking
the symmetry of the system, and so may induce orientation of the
proposed ground state.  This remains to be tested.

We show that surface morphology and its commensurate charge
distribution have an impact on correlations in the 2DES.  
Further advancement in heterostructures designed to support electron
correlation effects must overcome this systematic disorder.


\end{multicols}


\begin{thebibliography}{10}

\bibitem{Tsui:82}
D.C. Tsui, H.L. Stormer, and A.C. Gossard, 
Phys. Rev. Lett. {\bf 48}, 1559 (1982).
\bibitem{Stormer:79}
H.L. Stormer, R. Dingle, A.C. Gossard, W. Wiegmann, M.O. Sturge,
Solid State Comm. {\bf 29}, 705 (1979). 
\bibitem{Pfeiffer:88}
L. Pfeiffer, K.W. West, H.L. Stormer, K.W. Baldwin,
Appl. Phys. Lett. {\bf 55}, 287 (1988).
\bibitem{Olle:98}
for overview see Composite Fermions,
edited by O. Heinonen, (World Scientific, 1998).
\bibitem{Du:99}
R.R. Du, D.C. Tsui, H.L. Stormer, L.N. Pfeiffer, K.W. West,
Solid State Commun. {\bf 109}, 389 (1999).
\bibitem{Lilly:99}
M.P. Lilly, K.B. Cooper, J.P. Eisenstein, L.N. Pfeiffer, K.W. West,
Phys. Rev. Lett. {\bf 82}, 394 (1999).
\bibitem{Koulakov:96}
A.A. Koulakov, M.M. Fogler, B.I. Shklovskii,
Phys. Rev. B{\bf 76}, 499 (1996) and
R. Moessner, J. Chalker,
Phys. Rev. B{\bf 54}, 5006 (1996).
\bibitem{Willett:97}
R.L. Willett, K.W. West, L.N. Pfeiffer,
Phys. Rev. Lett. {\bf 78}, 4478 (1997).
\bibitem{Johnson:94}
M.D. Johnson, C. Orme, A.W. Hunt, D. Graff, J. Sudijono, L.M. Sander, B.G. Orr,
Phys. Rev. Lett. {\bf 72}, 116 (1994).
\bibitem{Fleming:80}
R.M. Fleming, D.B. McWhan, A.C. Gossard, W. Wiegmann, and R.A. Logan,
J. Appl. Phys. {\bf 51}, 357 (1980).
\bibitem{Alexandre:85}
F. Alexandre, L. Goldstein, G. Leroux, M. C. Joncour, H. Thibierge, E.V.K. Rao,
J. Vac. Sci. Technol. B {\bf 3}, 950 (1985).
\bibitem{Smet:98}
J.H. Smet, K. von Klitzing, D. Weiss, W. Wegscheider,
Phys. Rev. Lett.{\bf 80}, 4538 (1998).
\bibitem{MacDonald:00}
A.H. MacDonald, M.P.A. Fisher,
Phys. Rev. B {\bf 61}, 5724 (2000).
\bibitem{vonoppen:98}
Felix von Oppen, Ady Stern, and Bertrand I. Halperin,
Phys. Rev. Lett. {\bf 80}, 4494 (1998).
\bibitem{Willett:99}
R.L. Willett, K.W. West, L.N. Pfeiffer,
Phys. Rev. Lett. {\bf 83}, 2624 (1999).
\bibitem{Muller:92}
J.E. Muller,
Phys. Rev. Lett. {\bf 68}, 385 (1992).



\end{thebibliography}
\end{document}